\documentclass[preprint,3p,twocolumn]{elsarticle}
\usepackage{scrextend}
\usepackage{graphicx, epsfig,amssymb} 
\usepackage{amsmath, amsfonts}
\usepackage{bm}
\usepackage[breaklinks]{hyperref}
\usepackage{color}
\usepackage{enumerate}
\usepackage{amsmath}
\usepackage{cuted}
\usepackage{aas_macros}
\usepackage{capt-of}
\usepackage{adjustbox}

\newcommand{\DM}{\mathrm{DM}}
\newcommand{\DDDM}{\mathrm{DDDM}}
\newcommand{\reff}{\mathrm{ref}}

\newcommand{\di}{\mathrm{d}}

\begin{document}
\begin{frontmatter}
\title{
\vspace{-1.4cm}
\begin{flushright}
\small CERN-TH-2017-177
\end{flushright}
\vspace{0.8cm}
Binary pulsars as probes of a Galactic dark matter disk}

\author{Andrea Caputo\fnref{fn1}}\ead{andrea.caputo@uv.es}\address{Theoretical Physics Department, CERN, CH-1211 Geneva 23, Switzerland}\fntext[fn1]{Permanent address: Instituto de F\'{i}sica Corpuscular, Universidad de Valencia and CSIC, Edificio Institutos Investigaci\'{o}n, Catedr\'{a}tico Jos\'{e} Beltr\'{a}n 2, 46980 Spain}

\author{Jes\'us Zavala}\ead{jzavala@hi.is}\address{Center for Astrophysics and Cosmology, Science Institute, University of Iceland, Dunhagi 5, 107 Reykjavik, Iceland}

\author{Diego Blas}\ead{diego.blas@cern.ch}
\address{Theoretical Physics Department, CERN, CH-1211 Geneva 23, Switzerland}

\begin{abstract} 
As a binary pulsar moves through a wind of dark matter particles, the resulting dynamical friction modifies the binary's orbit.  We study  this effect for the double disk dark matter (DDDM) scenario, where a fraction of the dark matter is dissipative and  settles into a thin disk. For binaries within the dark disk, this effect is enhanced due to the higher dark matter density and lower velocity dispersion of the dark disk, 
and due to its  co-rotation with the baryonic disk.
We estimate the effect and compare it with observations for two different limits in the Knudsen number ($Kn$). 
First, in the case where DDDM is effectively collisionless within the characteristic scale of the binary ($Kn\gg1$) and ignoring the possible interaction between the pair of dark matter wakes. Second,  in the fully collisional case ($Kn\ll1$), where a fluid description can be adopted and the interaction of the  pair of wakes is taken into account. We find that the change in the orbital period is of the same order of magnitude in both limits. 
A comparison with observations reveals good prospects to  probe currently allowed DDDM models
with  timing data from binary pulsars in the near future. 
We finally comment on the possibility of extending the analysis to the intermediate (rarefied gas) case with $Kn\sim1$.  
\end{abstract}
\end{frontmatter}


\section{Introduction}
Unravelling the nature of dark matter (DM) is among the most fundamental frontiers of modern physics. Despite the growing body of gravitational evidence pointing to this new form of matter, its properties as a particle remain elusive. Given the lack of
any convincing signature of non-gravitational DM interactions, the gravitational effect of DM on ordinary matter remains the only {\it direct} source of observational information about  DM properties. 

Among the many gravitational DM probes, binary pulsars have been recently proposed as a novel way of deriving model-independent upper bounds on the dark-matter density at different distances from the Galactic center \cite{Pani:2015}. The presence of DM modifies the binary's orbit due to dynamical friction as the binary moves through the DM ambient medium. Given the extraordinary precision achieved in the measurement of the orbital properties of pulsar binaries, it was concluded in  \cite{Pani:2015} that these objects could be used to put constraints on the central DM density of the Milky Way. Isolated and binary pulsars have also been suggested 
to explore the scenarios of ultra-light DM candidates, see \cite{Khmelnitsky:2013lxt,Porayko:2014rfa,Blas:2016ddr,DeMartino:2017qsa}. 

In this paper we argue that the potential of binary pulsars as precise probes of the DM distribution increases substantially if one assumes that a fraction of the DM in our galaxy is distributed in a thin disk coplanar and co-rotating with the luminous disk. Such a possibility is materialized in models such as the Partially Interacting DM (PIDM) scenario proposed by  \cite{Fan:2013,Fan:2013PRL}, where a small fraction of the DM can lose energy to collapse into a thin `dark disk' (in an analogous way as ordinary matter assembles into disk galaxies). This so-called Double-Disk-Dark-Matter (DDDM) scenario has a distinct prediction for the DM phase space distribution in the disk, characterized by a high density, co-rotation with the luminous disk, and small velocity dispersion. Despite its striking features, the possibility of such dark disk remains observationally allowed  \cite{Kramer:2016} and there are already different methods suggested for its detection\cite{Belo:2017,Lelli:2015}. 
In the following we argue that all these effects increase the dynamical friction induced in binary systems. Remarkably, 
the enhancement is such that the sensitivities of current measurements of binary pulsars near the galactic plane are reasonably close to probe currently allowed DDDM scenarios. 
Notice that our results may be applied to any model where DM is expected to generate a disk with properties similar to those of the DDDM scenario,  e.g. \cite{DAmico:2017lqj,Foot:2014uba,Arkani-Hamed:2016rle,Buckley:2017ttd}. 

Our work is organized as follows: in Sec.~\ref{sec:dddm_model} we summarize the features of the DDDM model relevant for our analysis. Sec.~\ref{sec:collisionless} describes the dynamical friction for the collisionless DM case and ignoring the interaction of the wakes. We also compare our predictions with observations in that section under the assumption of co-rotation of the DDDM dark disk with the center of mass of the binary. In Sec.~\ref{sec:fluid} we extend this analysis to
the case of large Knudsen numbers and including the wakes' interactions.  We conclude and present our outlook in Sec.~\ref{sec:summary}.

\section{DDDM model}\label{sec:dddm_model}

The DDDM scenario discussed in \cite{Fan:2013,Fan:2013PRL} proposes the existence of a subdominant component of the DM sector that has dissipative dynamics.  It consists of a massless $U(1)_D$ gauge boson, with fine structure $\alpha_D$ and interactions with two new fields: a heavy fermion $X$ and a light fermion $C$, with opposite charges $q_x=1=-q_C$ under $U(1)_D$.  The thermal history of this scenario is described in detail in  \cite{Fan:2013}. In the following, we only describe the elements of the formation of a dark disk relevant to our purposes.

The formation of a dark disk occurs in an analogous way to baryonic matter assembling into galactic disks. The fraction of DM that is dissipative would fall into the gravitational well of the protohalos predominantly made of the non-interacting DM fraction, which has acquired angular momentum through tidal torques with the surrounding environment. At the time of accretion, part of the DDDM might be in the form of atomic-like states made of heavy and light DM. These dark atoms will become fully ionized due to shock heating during the virialization of the halo leaving a population of free $X$ and $C$ fermions. This dark plasma cools through Compton scattering and Bremsstrhalung off background dark photons \cite{Rosenberg:2017}. Since the plasma carries angular momentum it will form a rotationally supported dark disk as it dissipates energy and collapses due to gravity. Torques between the dark and baryonic disks will tend to align them into a steady configuration. In the following, we assume that this state has been reached with both discs being aligned and co-rotating\footnote{Even when this condition is not fully satisfied, a systematic study of 
all binary systems may allow to exclude the presence of a dark disk.}.

The cooling process is important in our discussion because it sets the {\it vertical} velocity dispersion of the disk, which is approximately given by the final temperature of the cooled plasma: $\sigma_z^2\approx T_{\rm cooled}/m_X$, where $m_X$ is the mass of the heavy fermion $X$. In  \cite{Fan:2013}, assuming the ionization fraction to be between $1\%$ and $10\%$, the authors estimate $T_{\rm cooled}\sim(0.02-0.2)B_{XC}$, where $B_{XC}=\alpha_D^2m_C/2$ is the binding energy of the ground state of the dark atom and $m_C$ is the mass of the light fermion $C$. For instance, for $m_X=100\,{\rm GeV}$, $m_C=1$\,{\rm MeV} and $\alpha_D=0.1$  this estimate results in $\sigma_z\sim9.5$~km/s for the region of parameter space where cooling is efficient (see Fig.~5 and Fig.~7 of \cite{Fan:2013}). This dispersion is at least an order of magnitude smaller than the typical one in the solar neighborhood for a Milky-Way-size halo formed of collisionless DM particles ($\sigma_{\rm 1D}\sim100-130$~km/s; e.g. \cite{Kuhlen2010}), where the DM orbits are no longer circular. As we will see below, a low velocity dispersion plays a key role in the constraining power of binary pulsars in the DDDM scenario.

Simulations of the formation and evolution of a DDDM disk have not been performed yet, and thus its final distribution remains an analytical approximation, which has been modelled as an isothermal sheet \cite{Binney:2008}, with a density given by:
\begin{align}\label{eq_rho}
\begin{split}
\rho(R,z)=\dfrac{\epsilon_{\rm disk} M_{\rm DM}^{\rm gal}}{8 \pi R_d^2 z_d}e^{-R/R_d}{\rm sech}^2(z/2z_d),
\end{split}
\end{align}
where $R$ is the radial distance in the plane of the disk and $z$ indicates the height above the disk midplane; $R_d$ and $z_d$ are the scale lengths of the disk in the radial and vertical directions, respectively. The mass fraction of DM in the halo of the Milky Way that could be in the form of a disk is denoted by  $\epsilon_{\rm disk}\equiv M_{\rm DDDM}^{\rm disk}/M_{\rm DM}^{\rm gal}$. The parameters of the disk, regardless of the PIDM nature, are constrained by the kinematics of the stars in the solar neighborhood ($R=R_\odot$). In particular, the surface density of the disk below a certain height $z_0$,
\begin{align}\label{sigma_d}
\begin{split}
		\Sigma_{\rm disk}(R_\odot,\left | z \right | < z_0)\equiv\int_{-z_0}^{z_0} \rho (R_\odot,z)\, \di z \, , \\ 
\end{split}
\end{align}
is strongly constrained by local stellar kinematics for $z_0\lesssim1$~ kpc. If the scale length $R_d$ is assumed to be similar to that of the baryonic disk of the Milky Way, $R_d\sim3$~kpc, and since the DDDM will be thin ($z_d\ll z_0\sim1$~kpc), then Eq.~\eqref{sigma_d} reduces to:
\begin{align}\label{sigma_d2}
\begin{split}
\Sigma_{\rm disk}(R_\odot, \left | z \right | < 1~{\rm kpc})\sim\dfrac{\epsilon_{\rm disk} M_{DM}^{gal}}{2 \pi R_d^2}e^{-R_\odot/R_d}.
\end{split}
\end{align}

Given that we are assuming $R_d\sim3$~kpc and $R_\odot\sim 8$~kpc, and the total halo mass of the Milky Way is $\sim10^{12}$~$M_\odot$, then a constraint on Eq.~\eqref{sigma_d2} translates into a constrain on $\epsilon_{\rm disk}$. A compilation of current bounds on the local surface and volume densities of total matter and visible matter is given in  \cite{Kramer:2016} (see their Table 1). For instance, using the results from  \cite{Bovy:2013raa}, the surface density of non-baryonic matter is constrained to be: 
\begin{align}
\begin{split}
\Sigma_{\rm dark}(R_\odot, \left| z \right| < 1.1 {\rm kpc})= 30\pm4\dfrac{M_{\odot}}{{\rm pc}^2},
\end{split}
\end{align}
which implies $\epsilon_{\rm disk}<0.025$, i.e., only a few percent of the total halo mass can be in the form of DDDM. An updated analysis on the constraints on a dark disk by  \cite{Kramer:2016}  concluded that depending on the method (standard static or one 
including non-equilibrium features in the tracer stars), the upper bound (95\% confidence interval) on $\Sigma_{\rm disk}$ lies between
$3-13$~$M_\odot$/pc$^2$ for a thin disk ($z_d=10$~pc), while for a thick disk ($z_d=100$~pc), the upper bound changes to $7-32$~$M_\odot$/pc$^2$ (the bound for the non-equilibrium method was extracted from Fig.~10 of  \cite{Kramer:2016}). A thin dark disk model with these characteristics has been invoked to potentially explain the apparent periodicity of comet impacts on Earth \cite{Randall:2014}. One finds:
\begin{eqnarray}\label{eq:rho_sigma}
\rho_0^{\rm DDDM}&\equiv&\rho(R_\odot,z=0)= \frac{\Sigma_{\rm disk}(R_\odot)}{4z_d},\\
\sigma^2_z&=&8\pi G \rho_0^{\rm DDDM} z_d^2,
\end{eqnarray}
for the midplane density and the vertical velocity dispersion of the disk, respectively. Thus, the stellar kinematic upper bounds on the dark disk, imply for a thick disk ($z_d=100$~pc):
\begin{eqnarray}\label{eq:bound_thick}
\rho_0^{\rm DDDM}(\rm thick)&\lesssim&\,3\,{\rm GeV}/{\rm cm}^{3},\\\nonumber
\sigma_z(\rm thick)&\lesssim&\,9\,{\rm km/s},
\end{eqnarray}
while for a thin disk ($z_d=10$~pc):
\begin{eqnarray}\label{eq:bound_thin}
\rho_0^{\rm DDDM}(\rm thin)&\lesssim&\,12\,{\rm GeV}/{\rm cm}^{3},\\\nonumber
\sigma_z(\rm thin)&\lesssim&\,2\,{\rm km/s}.
\end{eqnarray}
This is the range of values of interest for our purposes.

\section{Effect for collisionless DDDM and non-interacting wake pair}\label{sec:collisionless}

In this section we follow closely the approach of  \cite{Pani:2015} and study a binary system with masses $m_{1}$ and $m_{2}$ moving through a distribution of {\it collisionless} DM with a local constant density $\rho_{\rm DM}$. This assumption is justified as long as the local
mean free path for self-interactions is much larger than the characteristic scale of the DM wakes induced by dynamical friction in the orbital time of the binary.  As done in  \cite{Pani:2015}, we ignore the possible interaction of the wakes.
We comment on the collisional regime and the expected (small) influence of the wakes' interaction in Section \ref{sec:fluid}.

Neglecting relativistic effects\footnote{In particular, we neglect gravitational waves (GW) emission. This assumption will always hold in the non-relativistic binaries of interest for our purposes since large periods are tipically needed to enhance the drag force signal. For a discussion on GW emission see e.g. \cite{Gomez:2017wsw}.}, the equations of motion for the two bodies are:
\begin{align}
\begin{split}\label{eq:DMbin}
		m_{i}\ddot{\bf r}_{i}=\pm\dfrac{Gm_{1}m_{2}}{r^3}\mathbf{r}+\mathbf{F}_i^{\mathrm{DM}},
\end{split}
\end{align}
where $i=1,2$ refers to the two members of the binary, each with a position $\mathbf{r}_i$ and mass $m_i$, $\mathbf{r}=\mathbf{r}_2-\mathbf{r}_1$ is their relative position, and $\mathbf{F}_i^{\mathrm{DM}}$ is the drag force (dynamical friction) due to DM acting on the $i$-th body. To compute this force, we first refer to the situation of a body of mass $M$, moving with velocity $\mathbf{v}$ through a homogeneous density distribution of collisionless particles with mass $m_{\rm DM}\ll M$ and velocity distribution $f(\mathbf{u})$.

 The problem can be analyzed as a set of sequential encounters of the object $M$ with particles randomly taken from the distribution $f(\mathbf{u})$, over an interval of time much shorter than the time scale for variations in the velocity $\mathbf{v}$, and much longer than the interaction time. By symmetry arguments, the variation of $\mathbf{v}$ perpendicular to the direction of motion ($\Delta\,v_{\bot}$) vanishes, while the change in the parallel component is given by \cite{Gould:1991}: 
\begin{align}\label{df_0}
\begin{split}
		\dfrac{\di v}{\di t}&=-\dfrac{4\pi G^2Mm_{\rm DM}}{v^2}\\
		&\left[\int_{0}^{v} f(u)\left[\ln \Lambda+\ln \left(\dfrac{v^2-u^2}{v^2}\right)\right]\, \di^3u\right. \ \\ 
		& +\left.\int_{v}^{\infty}f(u)\left[-\dfrac{2v}{u}+\ln \left(\dfrac{u+v}{u-v}\right)\right]\, \di^3u\right],
\end{split}
\end{align}
where $\ln \Lambda \equiv\lambda = \ln (b_{max}v^2/GM)$, is the Coulomb logarithm. The scale length $b_{max}$ is the characteristic size of the medium, while $GM/v^2$ is the typical radius of the sphere of gravitational influence of the orbiting body. The choices of the parameters are somewhat arbitrary. In the following we choose them such that $\Lambda \gg1$ and in particular $\lambda = 20^{+10}_{-10}$, to make a direct comparison with the results of \cite{Pani:2015} in the standard halo scenario.

In the limit where the velocity dispersion of the medium is much larger than the velocity of the dragged object ($\sigma\gg v$), the equation above greatly simplifies and reduces to two contributions roughly of the same order \cite{Pani:2015,Gould:1991}. In this limit, for an isotropic Maxwellian distribution, Eq.~\eqref{df_0} reduces to the well known  
Chandrasekhar's formula \cite{Chandra:1943,Chandra:1949}:
\begin{align}
\begin{split}
\label{eq:Chandra}
		\dfrac{\di v}{\di t}=-\dfrac{4\pi G^2M \rho_{\rm DM}}{v^2}\lambda\left[{\rm erf}\,(x)-\dfrac{2x}{\sqrt{\pi}}\exp(-x^2)\right],
\end{split}
\end{align}
in which we have defined $x=v/\sqrt{2}\sigma$ and used the Gauss error function ${\rm erf}\,(x)$.
In \cite{Pani:2015} the constraints on the DM density from binary pulsars are computed in this limit ($\sigma\gg v$), using the previous formula\footnote{The validity of Eq.~\eqref{eq:Chandra} and the comparison to the general equation \eqref{df_0} are discussed in Appendix A of \cite{Pani:2015}. In the limit $\sigma\gg v$ they are indeed a very good approximation as can be seen in their Figure 7.}. Since we are mostly interested in the regime $\sigma\lesssim v$, we use the general equation \eqref{df_0}.

The previous effect represents a total deceleration of the object with mass $M$. For a binary pulsar,  we will be mainly interested in the change in the period of the binary.
Since in this section we neglect the possible interaction of one component with its companion's wake, we can use the superposition of forces $\mathbf{F}_i^{\mathrm{DM}}$ for $i=1,2$. For {\it collisionless} DM, this approximation holds when the period of the binary $P_b$ is larger than the typical time of dispersion of the wake \cite{Bekenstein:1990}: 
\begin{align}
\begin{split}
		P_b\gg\dfrac{Gm_i}{\sigma^3}.
\end{split}
\end{align}
In the case of a dark disk, the low value of $\sigma$ implies $P_b\lesssim Gm_i/\sigma^3$ for typical binary periods of $10-100$ days and $\sigma \sim 1-10$~km/s. As a result, one should 
{\it in principle} include the interaction of the wakes in the analysis. This study is currently missing in the literature and
we leave its detailed treatment for the future. We anticipate, however,  that the impact of this interaction that one finds in the case of a gaseous medium  \cite{Kim:2007,Kim:2008} (see also Section~\ref{sec:fluid})  suggests that the calculation that follows captures the correct order of magnitude of the effect.  
 
\subsection{Osculating orbits and orbital period variations caused by DDDM}

\begin{figure}
       \centering
     \includegraphics[width=0.4\textwidth]{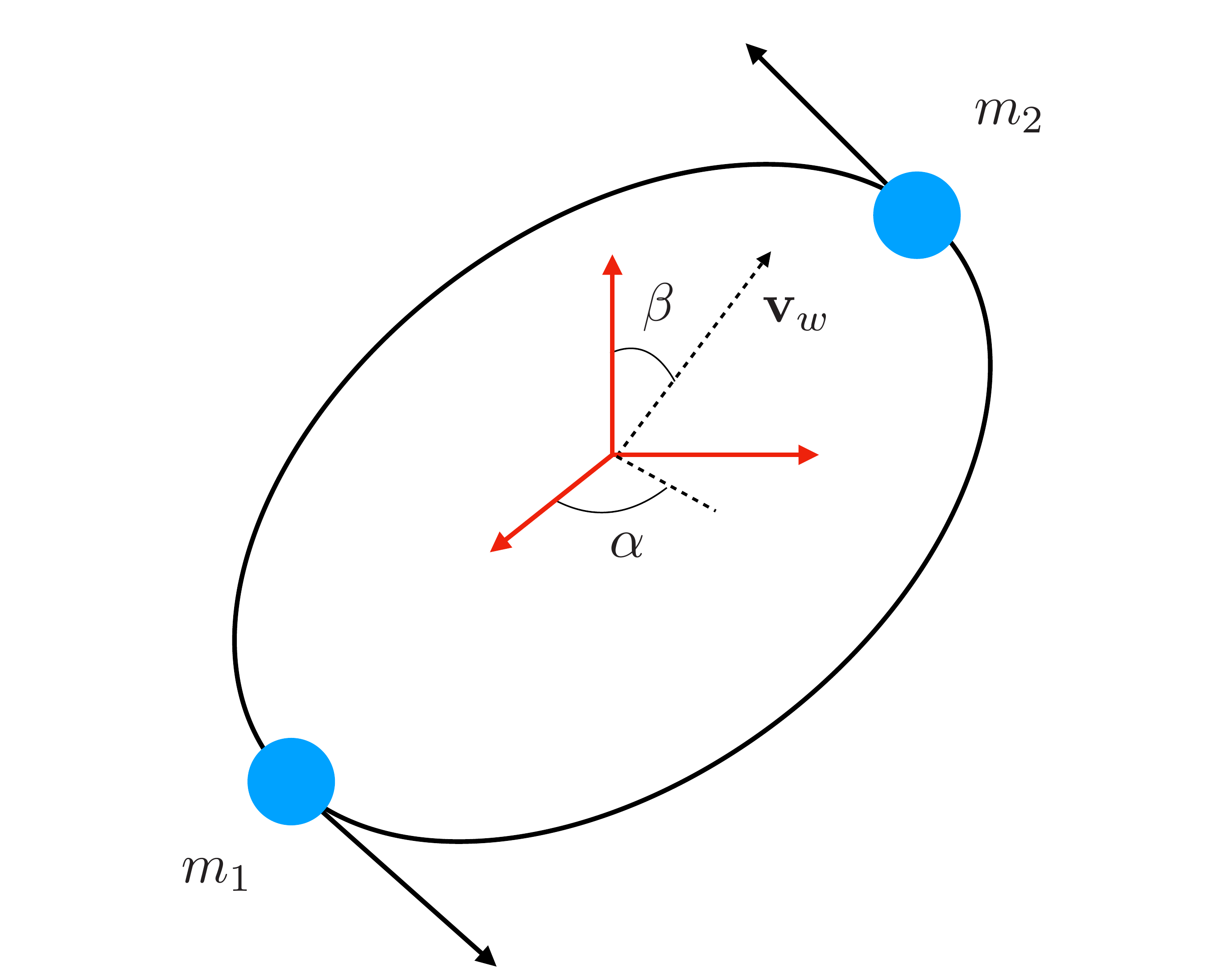}
        \caption{Scheme of a generic binary system in the Galactic reference frame. This is the general situation in which a DM wind could have a net velocity relative to the center of mass of the binary: $\mathbf{v}_{w}=v_{w}({\rm cos} \alpha\, {\rm sin} \beta, {\rm sin} \alpha\, {\rm sin} \beta, {\rm cos} \beta)$.}
 \label{fig:sketch}
 \end{figure}

We introduce the position vector for the center of mass $\mathbf{R}=(m_1\mathbf{r_1}+m_2\mathbf{r_2})/M$ with $M=m_1+m_2$ and rewrite the drag force as $\mathbf{F}_i^{\mathrm{DM}}=-Ab_i(m_i^2/M)\tilde{\mathbf{v}}_i$, where $A=4\pi\rho_{\rm DM}G^2M$, and:
\begin{align}\label{eq:bi}
\begin{split}
	&b_i=\dfrac{1}{\tilde{v_i}^3}\left[\int_{0}^{\tilde{v_i}} f(u)\left[\ln \Lambda+\ln \left(\dfrac{\tilde{v_i}^2-u^2}{\tilde{v_i}^2}\right)\right]\, \di^3u\right.\\
	&~~~~~~~\left.+\int_{\tilde{v_i}}^{\infty}f(u)\left[-\dfrac{\tilde{2v_i}}{u}+\ln \left(\dfrac{u+\tilde{v_i}}{u-\tilde{v_i}}\right)\right]\, \di^3u\right],
	\end{split}
\end{align}
with $\tilde{\mathbf{v}}_i=\dot{\mathbf{r}}_i+\mathbf{v}_w$ being the velocity of the $i$-th companion relative to the DM wind, see Fig.~\ref{fig:sketch}  (adapted from \cite{Pani:2015}). We  assume the net velocity relative to the center of mass of the binary to be negligible 
 $\mathbf{v}_w\approx0$ because in the simplest dark disk scenario the dark disk co-rotates with the galactic disk. It is important to notice that this assumption is not 
 always satisfied for binary pulsars.  In particular, the velocity kick followed by the supernova explosion is of order\footnote{We are grateful to Lijing Shao for pointing out this 
 to us.}
 $\gtrsim 50\ \mathrm{km/s}$ (e.g. \cite{Wang:2006}; notice that current uncertainties in kick velocities are very large). However, with a sufficient number of systems one expects that some of them will have relative velocities $v_w \lesssim \sigma$, which is the relevant limit for our conclusions, cf.~Fig.~\ref{fig:pb_vwind}. Thus, to maximize the possible effect of dynamical friction, we assume that  the DDDM particles and the center of mass of the binary  move in approximately the same circular orbit, with nearly identical velocities. 
Under this condition, the equations of motion reduce to:
\begin{eqnarray}
	\mathbf{\dot{v}}&=&-\dfrac{GM}{r^3}\mathbf{r}+a_1\eta\mathbf{v}+a_2\cdot\mathbf{V},\\
	\mathbf{\dot{V}}&=&a_2\eta\mathbf{v}+a_3\cdot\mathbf{V},
\end{eqnarray}
where $\mathbf{v}=\mathbf{\dot{r}}, \mathbf{V}=\mathbf{\dot{R}}$, and:
\begin{eqnarray}\label{eq:a1_a2}
	a_1&=&-A(b_1+b_2),\\\nonumber
	a_2&=&\dfrac{A}{2}\left[b_1 \Delta_++b_2\Delta_-\right],\\\nonumber
 	a_3&=&-\dfrac{A}{4}\left[b_1\Delta_+^2+b_2\Delta_-^2 \right],
\end{eqnarray}
 with $\Delta_\pm=\Delta\pm1$, $\Delta=\sqrt{1-4\eta}$, $\eta=\mu/M$ and $\mu$ is the reduced mass of the binary.
We treat the drag force as a perturbation and use the formalism of osculating orbits \cite{Poisson:2014} to make the perturbative analysis to first order. Following\cite{Pani:2015,Poisson:2014} and restricting to circular orbits for the binary for simplicity (a good approximations for the systems we refer to given their small eccentricities), we obtain an expression for the time derivative of the orbital period:
\begin{equation}\label{p_decay}
 	\dot{P}^{\mathrm{DM}}_b(t)=3P_b[a_1\eta-a_2\Gamma \, {\rm sin} \beta \,{\rm sin} (\Omega_0t- \alpha)],
\end{equation}
where $P_b$ is the period of the orbit, $\Gamma=v_{w}/v$, $\Omega_0$ is the orbital angular velocity and $\alpha$ and $\beta$ are the azimuthal and polar angles, respectively (see Fig.~\ref{fig:sketch}).  If $v_w=0$, then $\Gamma=0$, and since we are assuming a circular orbit, the relation $(GM\Omega_0)^{1/3}=\vert\mathbf{v}\vert\equiv v$ holds, where $v$ is the relative velocity of the binary members. In this case, Eq.(\ref{p_decay}) greatly simplifies.

\subsection{Changes in the period  in the DDDM scenario}

As a benchmark, let us consider a binary system with  the following orbital parameters: $P_b=100$~days, $m_1=1.3$~$M_{\odot}$ and $m_2=0.3$~$M_{\odot}$. In the ordinary DM scenario, typical values for the other relevant quantities (at the solar circle) are $v_{w}\sim220$~km/s,   $\sigma\sim150$ km/s and $\rho_{\odot}^{DM}=0.3\,$GeV$\,$cm$^{-3}$. Using Eq.~\eqref{p_decay} in this case, the secular change of the orbital period results in: 
\begin{equation}
\label{eq:Pani}
\vert\dot{P_b}^{\DM}\vert=6.6\cdot10^{-19},
\end{equation}
which is too small to be probed by current measurements. Furthermore, it seems challenging to identify 
such small effect in the future given the typical uncertainties in other known mechanisms that affect $\dot P_b$ \cite{Pani:2015,Damour:1991,lorimer2005handbook}. 

The expectations drastically change when one considers the DDDM scenario, and $v_{w}\sim0$. In this case, for a binary system with the same characteristics as above (located at the solar circle), with $z=10$~pc, immersed in a thin dark disk with $ z_d=10$~pc and thus $\sigma=1.8$~km/s (given Eq.~\eqref{eq:rho_sigma}) one obtains:
\begin{equation}
\vert\dot{P_b}^\DDDM\vert=4.2\cdot10^{-13} 
\end{equation}
while considering a thicker dark disk, with $z_d=100$~pc ($\sigma=9$~km/s), one would get:
\begin{equation}
\vert\dot{P_b}^\DDDM\vert=1.42\cdot10^{-14}.
\end{equation}
These numbers may be within the reach of current  or future observations  (see below). The dependence of $\vert\dot{P_b}^\DDDM\vert$ on $P_b$ is shown
in Fig.~\ref{fig:data_comparison} below for the two dark disk cases we have discussed. At small periods, velocities are large and since the dispersion is very small, the term that dominates in Eq.~\eqref{eq:bi} is the first integral, i.e. DM particles with velocities smaller than $v_i$ dominate the drag. This can be seen more clearly in Eqs. (A.2-A.5) of \cite{Pani:2015}. The result is that roughly $b_i\propto1/v^3$ (since that integral does not have a very strong dependence on $v_i$). This  means that Eq.~\eqref{p_decay} goes as $1/v^4\sim P_b^4$, which is roughly the scaling we see in Fig.~\ref{fig:data_comparison}. At large periods, the circular velocities are smaller and the second integral in  Eq.~\eqref{eq:bi} starts to become important. The latter  has a stronger dependence on $v$ which makes the scaling of $\vert\dot{P_b}^\DDDM\vert$ with $P_b$ shallower.
\begin{figure}
       \centering
\includegraphics[width=\linewidth]{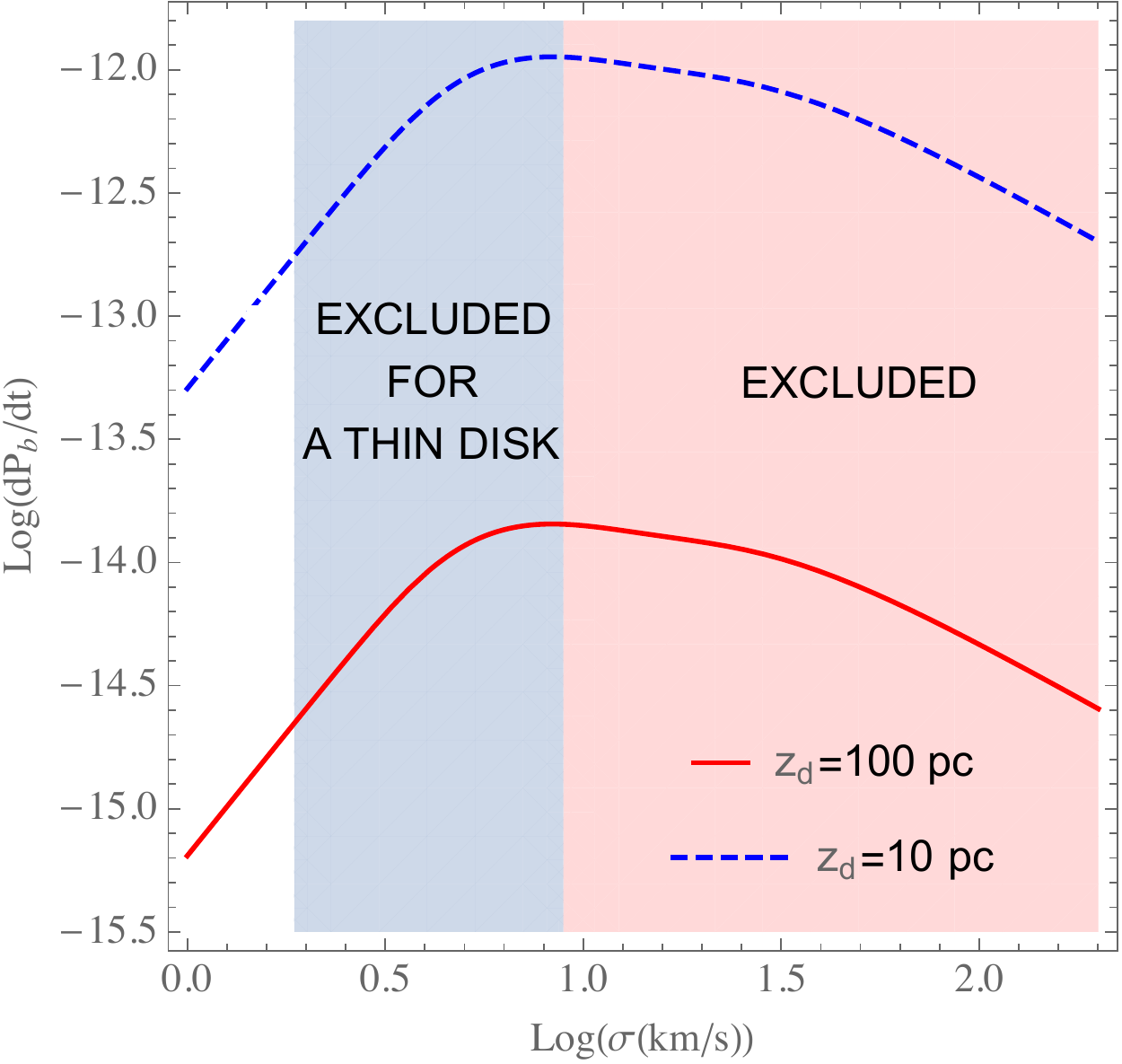}
\caption{$\dot{P}^\DDDM_b$ as a function of $\sigma$ in the DDDM scenario, for both a thin (red-thick line) and thick (blue-dashed line) disk, for a system with $P_b=100$ days, $m_1=1.3$~$M_{\odot}$, $m_2=0.3$~$M_{\odot}$ and $v_{w}=0$. The red shaded region ($\sigma>9$~km/s) corresponds to the region excluded for the thick dark disk, while the grey shaded region extends the exclusion region for the thin disc, both required by constraints on the DDDM density (see Sec. \ref{sec:dddm_model}). }
\label{fig:pb_sigma}
\end{figure}

Let us now discuss the dependence of the previous results on $\sigma$. This is shown in Fig.~\ref{fig:pb_sigma} for the DDDM case and  $P_b=100$ days, where we see that the dependence is not as strong as in the ordinary case\footnote{For instance, in the ordinary case changing $\sigma$ from $200$~km/s to $50$~km/s for the system under study strengthens the signal by almost two orders of magnitudes \cite{Pani:2015}.}. This is because decreasing $\sigma$ also leads to a lower density (Eq.~\eqref{eq:rho_sigma}) and  $\dot{P}^\DDDM_b$ grows linearly with the density ($a_1$ and $a_2$ are proportional to $A$ in Eq.~\eqref{eq:a1_a2}, and $A\propto\rho_{DM}$). In fact, at the low $\sigma$ end, $\dot{P}^\DDDM_b\propto\rho_{DM}\propto\sigma^2$.  In the intermediate regime $v_{w}\ll\sigma\leq v_{orb}$, the inversely proportional dependence of $\dot{P}^\DDDM_b$ on $\sigma$ (clearly manifest in \cite{Pani:2015}) starts to become important and thus, $\dot{P}_b$ peaks at around $\sigma \sim 7-8\, {\rm km/s}$ followed by a sharp decline in the high $\sigma$ end as in the ordinary case. In the regime, $\sigma\gg v_{w},v_{orb}$, we are in the limit discussed in section C.1 of \cite{Pani:2015} and $\dot{P}_b^\DDDM\sim\rho_{DM}/\sigma^3\propto1/\sigma$. 

\begin{figure}
       \centering
  \includegraphics[height=8.5cm,width=8.5cm,keepaspectratio]{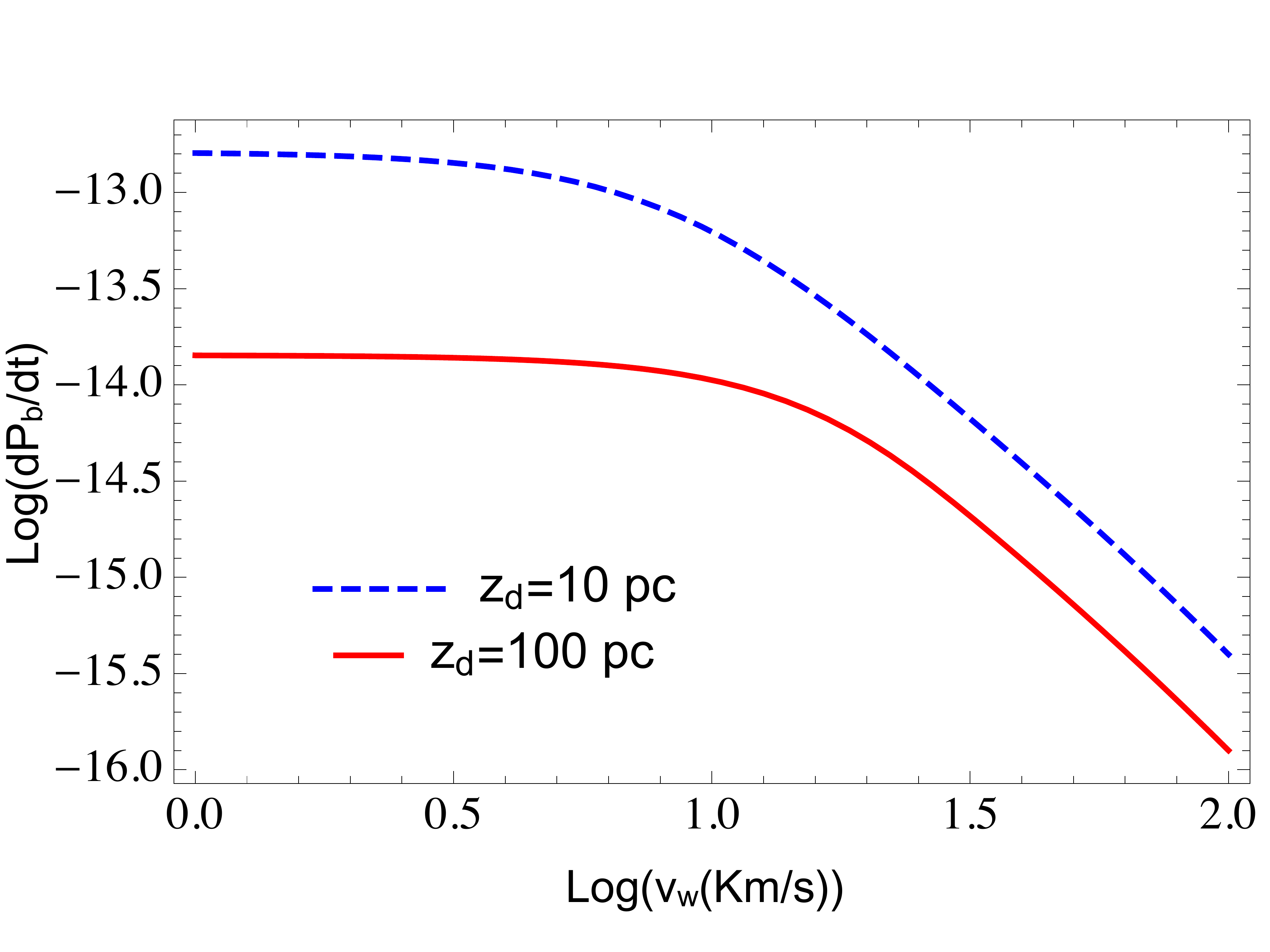}
\caption{$\dot{P}^\DDDM_b$ as a function of $v_{w}$ in the DDDM scenario with $P_b=100\,{\rm days}$, $m_1=1.3$~$M_{\odot}$, $m_2=0.3$~$M_{\odot}$.  We show both cases $z_d = 10$~pc (thin disk; blue-dashed line) and $z_d=100$~pc (thick disk; red-solid line), with corresponding dispersions $\sigma=1.8\;{\rm km/s}$ and $\sigma=9\;{\rm km/s}$. These models are at the limit on current constraints on the DDDM scenario. }
\label{fig:pb_vwind}
\end{figure}

Another significant feature of the enhancement of the signal in the presence of the dark disk is the assumed low value for $v_{{w}}$. As we discussed, this approximation is relevant given the co-rotation of the dark and baryonic disks and the fact that some binary systems may satisfy $v_w \lesssim \sigma$. As shown in Fig.~\ref{fig:pb_vwind},  larger values of $v_{w}$ lead to a smaller signal. Moreover, the dependence of $\dot{P}^\DDDM_{b}$ on $v_{w}$ becomes more relevant for $v_{w}\gg\sigma$. The two curves in Fig.~\ref{fig:pb_vwind} are flat for low values of $v_{w}$, comparable to $\sigma$, but when $v_{w}\geq\sigma$ the signal decreases rapidly. This happens for larger values of $v_{w}$ in the case of $z_d=100$~pc since $\sigma$ is larger. This behaviour can also be seen in Fig.~2 of  \cite{Pani:2015}.

The co-rotation with the baryonic disk is a common characteristic for all possible dark disks; in fact, even in the case in which a dark disk formed from collisionless DM accreting onto a stellar disk, it would co-rotate with the baryonic one. However, if we consider a typical situation for an ``ordinary" dark disk \cite{Purcell:2009}, we may have $v_{w}\sim\,0$, but crucially a dispersion similar to that of stars $\sigma\sim\,100$ km/s. Also, the dark disk in these cases has a density $\lesssim20\%$ of the host halo density at the solar circle. Therefore, the signal would be several orders of magnitude less relevant than in the DDDM case. For the binary with the characteristics we have been considering: 
\begin{equation}
\dot{P}_b^{\mathrm ODDM}\sim-7.5\cdot10^{-19},
\end{equation}
which is very similar to the effect from the smooth halo we estimated above \eqref{eq:Pani}. Thus, 
the method proposed is only efficient to constrain disks similar to those appearing in the DDDM scenario. 

\begin{figure}
       \centering
      \includegraphics[height=8.5cm,width=8.5cm,keepaspectratio]{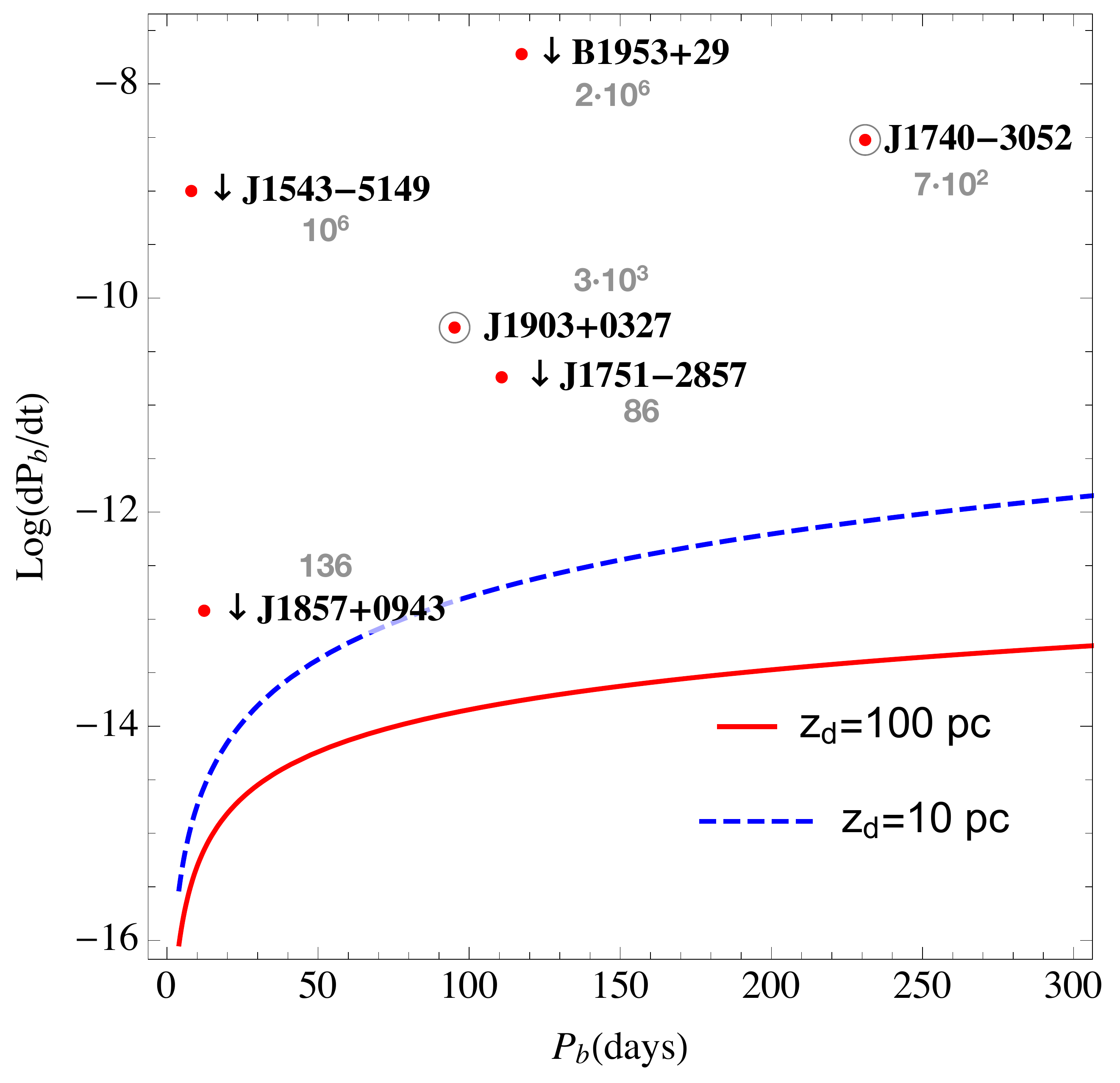}
\caption{Comparison between the $\dot{P_b}$ measured (or constrained) for most promising binaries and the predictions in the DDDM scenario for a binary with $m_1=1.3$~$M_{\odot}$, $m_2=0.3$~$M_{\odot}$. The predictions for $\dot P_b^\DDDM$ for a thin (thick) disk with $z_d=10(100)$~pc, and $\rho_0^{\rm DDDM}=\,12(3)\,{\rm GeV}\,{\rm cm}^{-3}$ are shown by a blue-dashed (red-solid) line and assuming $v_w \lesssim \sigma$. The downward arrows indicate experimental upper bounds on $\dot{P_b}$, while the circles with a central red dot refer to systems for which $\dot{P_b}$ has been directly measured. The grey values are the ratios of the predictions for the $\dot P^{\DDDM}_b$ of the system (maximal effect)  with respect to observational values/limits.}
\label{fig:data_comparison} 
\end{figure}

\subsection{Comparison with observations}

To compare our previous predictions with observations we have collected the  measured  value of (or upper limit  to) $\dot{P_b}$ for a number of binary pulsars near the galactic plane, and thus possibly immersed in a dark disk, and with periods  $P_b > 6\,$days.
 The theoretical prediction for $\dot{P}^\DDDM_b$ for a given system is computed by taking the coordinate of the binary above the plane ($z=z_{\rm sys}$), and assuming a dark disk with $z_d=z_{\rm sys}$. For the density, we take the largest upper bound on $\rho_0^{\rm DDDM}$ given by the kinematical constraints on $\Sigma_{\rm disk}(R_\odot)$ (see Eqs.~\eqref{eq:bound_thick}-\eqref{eq:bound_thin}; based on \cite{Kramer:2016}).  We further assume $v_w\lesssim \sigma\sim 10\ \mathrm{km/s}$. Although this approximation is not satisfied by some of the systems whose transverse velocities are large, we prefer to emphasize the case of almost co-rotation as it maximizes the constraining power of binary pulsars for the DDDM scenario. It is to be expected that a fraction of binary pulsars currently known, or to be detected in the future would actually satisfy this condition. 
 These numbers are collected in Table~\ref{table:systems} of the Appendix~\ref{ap:table}. 
   We find that, even though the signal in the DDDM scenario is much larger than in the standard one, the predicted values for most of the systems are several orders of magnitude away from experimental sensitivity. The main reason for this is that the effect is significant for relatively long periods, for
which an accurate measurement of $\dot P_b$ is more challenging. 
Nevertheless there are few interesting cases with relevant constraining power (gathered in Table~\ref{table:good}). We compare the constraints/measurements on $\dot P_b$ with the predictions for $\dot{P}^\DDDM_b$   in Fig.~\ref{fig:data_comparison}.   The observational
bounds to $\dot P_b$ are either directly taken from the references (given in the last column of Table~\ref{table:systems}, and denoted by $\reff$ in Table~\ref{table:good}) or estimated by considering the error in the 
determination on $P_b$ and the time of the observational campaign. 
 \begin{table}
\label{Table_short}
\begin{adjustbox}{width=0.48\textwidth,totalheight=\textheight,keepaspectratio}
\centering	
\begin{tabular}{lcccc}
\hline
 Name& $P_b$(days) &  $\dot{P}_b^\DDDM$ & $\dot{P}_b({\rm lim})<$ & Eccentricity\\
 \hline
       J1751-2857   & 110.7    & $-2.1\cdot10^{-13}$ & $1.8\cdot10^{-11}$ & $1.3\cdot 10^{-4}$\\ 
       J1857+0943   & 12.3     & $-9.76\cdot10^{-16}$ & $1.2\cdot10^{-13}$ & $2.2\cdot 10^{-5}$\\ 
    J1543-5149& 8.06  &$-6.47\cdot10^{-16}$&$10^{-9}$ & $2.2\cdot 10^{-5}$ \\
    J1740-3052& 231.03  &$-4.40\cdot10^{-12}$&$\reff$ $3\cdot10^{-9}$ & 0.58 \\
     J1903+0327& 95.17   &$-9.42\cdot10^{-15}$&$\reff$ $-33\cdot10^{-12}$ & 0.44  \\
    B1953+29& 117.35   &$-7.80\cdot10^{-15}$&$10^{-8}$ & $3.3\cdot 10^{-4}$\\
\hline
  \end{tabular}
\end{adjustbox}
\captionof{table}{A collection of promising binary pulsars to probe the DDDM scenario (Fig.~\ref{fig:data_comparison}). The columns are: binary period, expected maximal time variation in the binary period due to the presence of a dark disk, limits in the 
observed time variation of the binary's period (systems tagged with ``$\reff$'' have a measured bound on $\dot{P_b}$; see Table~\ref{table:systems}), and eccentricities.} 
\label{table:good}
\end{table}
The grey values next to each pulsar's name are the ratios of the predictions for the $\dot P^{\DDDM}_b$ of the system for the maximal effect  with respect to observational values/limits. The blue-dashed (red-solid) line shows the generic case of a binary ($m_1=1.3~M_\odot$, $m_1=0.3$~$M_\odot$) under the influence of the thin (thick) disk models that we have considered thus far,  with parameters at the upper boundary of current constraints. 

The situation summarized in Fig.~\ref{fig:data_comparison} can improve in the near future with the discovery of more systems with a combination of the following characteristics: (i) nearly co-rotating 
with the baryonic disk, (ii) closer to the Galactic Center (since the density is enhanced substantially relative to the Solar circle); (iii) with orbital periods $\gtrsim100$~days (given the  dependence of $\dot{P}^\DDDM_b$ with $P_b$ for smaller periods, see Fig.~\ref{fig:data_comparison}); and (iv) small orbital velocities, naturally connected to point (iii), given the strong inverse dependence of $\dot{P}^\DDDM_b$ on velocity (see Eq.~\eqref{eq:bi}). These desirable properties are not the best to obtain an optimal timing of the system, but these two aspects are not necessarily exclusive. Future surveys, such as SKA, will most likely add $\mathcal O(10^3)$ new systems \cite{Kramer:2015bea}, and hopefully some of them will have the right properties. 

We note however that even a system with a relatively short period can be quite constraining if a limit on $\dot{P_b}$ can be set   with  high precision. For instance, 
the binary system J1857+0943~\cite{Desvignes:2016}, whose period has been measured with 10 digits in precision during seven years, allows us to infer an upper bound: 
\begin{equation}
|\dot{P_b}|<1.2\cdot10^{-13},
\end{equation}
which is shown at the bottom left of Fig.~\ref{fig:data_comparison}. For this system, a thin dark disk in the DDDM scenario with
$z_d=z_{sys}$, would imply the following period variation caused by dynamical friction:
\begin{equation}
|\dot{P}^\DDDM_b|\sim8.6\cdot10^{-16},
\end{equation}
 J1751-2857~\cite{Desvignes:2016} is another promising system. Its period has been measured with 8 digits in precision for six years, which gives an upper bound:
\begin{equation}
|\dot{P_b}|<1.8\cdot10^{-11}.
\end{equation}
Since the period  of the binary is long ($P_b=110.7$~days) and its coordinates are suitable for a large density of the dark disk ($z_{\rm sys}=20$~pc and $R_{sys}\sim7.4$~kpc), the prediction for the impact of dynamical friction in the DDDM scenario is:
\begin{equation}
|\dot{P}^\DDDM_b|\sim2.1\cdot10^{-13}.
\end{equation}
a factor of $\sim86$ below the observational upper limit. 

We stress that the bounds derived for these systems are too optimistic given that the transverse velocity of the systems are $v_w\gtrsim v_T\sim 34\ \mathrm{km/s}$ and $v_w\gtrsim v_T\sim 44\ \mathrm{km/s}$, respectively\footnote{http://www.atnf.csiro.au/people/pulsar/psrcat/}. As we have highlighted before, these bounds are estimated under the assumption $v_w\lesssim \sigma$. Our purpose on presenting them here is to show that the timing of systems with the required properties can be achieved at the desirable level in the foreseeable future. Notice that if $v_w$ is indeed close to the transverse velocity of these binaries, then the bounds weaken by approximately a factor of $\sim10$, $\sim100$, respectively.

\section{Drag in a gaseous medium: interacting wake pair}\label{sec:fluid}

The previous analysis was done in the approximation of collisionless DM particles and ignoring the interaction between the
DM wakes. As we stressed, an analysis where the wakes' interaction is considered seems to be required to properly compute $\dot P_b^\DDDM$ in the systems 
we used for our analysis. To our knowledge, this is currently missing in the literature, but one can try to quantify its effect by treating DM as a gaseous medium, where the influence of the wakes has been fully analyzed in a different context. Independently of this motivation, the properties of DM are not well constrained and it is interesting to also constrain DM models beyond the collisionless paradigm, where a fluid approximation may be relevant.

Let us consider a binary system moving in an {\it inviscid gaseous medium}, which in our case is the DM plasma in the disk. For a model with an isothermal, self-gravitating Maxwellian distribution function, the speed of sound of the DM ``fluid" is proportional to the 1D velocity dispersion \cite{Binney:2008}: 
\begin{align}
\begin{split}
      c_s=\sqrt{\gamma}\sigma_{z},
\end{split}
\end{align}
where $\gamma$ is the adiabatic index ($\gamma=5/3$ for a monoatomic gas). 

We stress that in order to use this approach, we need to treat DM as a fluid.  
Dissipative self-interacting DM (like the one considered here) is often considered as a perfect fluid \cite{Clavijo:2014, Hu:2006} or even a plasma \cite{Clarke:2016}. Therefore this approximation seems to be accurate, especially in the inner regions of the disk, where the DM density is higher. 
To understand its validity more quantitatively, let us compare the mean free path $\ell$ of the medium with the characteristic length of the binary system (e.g. semi-major axis $a$). Following \cite{Rosenberg:2017} (Eq. (39), section 4.3), we estimate the mean free path for Rutherford scattering of charged DM particles as:
\begin{eqnarray}\label{eq:mean}
	\ell&\sim&10^{-3}~{\rm pc}\,\left(\frac{{\rm cm}^{-3}}{n_C}\right)\left(\frac{T}{10^6K}\right)^2\nonumber\\
	&&~~.\left(\frac{10^{-2}}{\alpha_D}\right)^2\left(\frac{21}{\ln\left(1+\frac{3T}{\alpha_D\cdot n_C^{1/3}}\right)}\right)
\end{eqnarray}
where $\alpha_D$ is the fine-structure constant for $U(1)_D$; $T$ is the temperature of the DM fluid: $kT/m_X\simeq\sigma_z$, and $n_C$ is the number density of light fermions, given by the ionization fraction $n_C/n$, where $n=n_{XC}+n_X$ with $n_{XC}$ the bound state number density and $n_X=n_C$ is the number density of heavy fermions. The ionization fraction is typically below $10\%$ for the interesting region of the DDDM parameter space where cooling is efficient \cite{Fan:2013}, and we have $\rho^{\DDDM}\sim m_X n$. For instance, if $\sigma_z=1$~km/s (which gives $T\sim$~eV), and for $n_C/n=0.1$, $\alpha_D=10^{-2}$, $m_X=100$~ GeV and $\rho^{\DDDM}=10$~GeV/cm$^3$, we have:
\begin{equation}
\ell\sim10^{-5}~{\rm pc},
\end{equation}
which is approximately 2 astronomical units. This  implies a Knudsen number $Kn=\ell/a\gtrsim1$, where $a$ is the major axis of the binary system, which is not truly in the (fully collisional) fluid regime $Kn\ll1$, but rather intermediate, corresponding to the case of a rarefied gas. 
However, given the dependence of Eq. \eqref{eq:mean} in, for instance $\alpha_D$, it is evident that a part of the interesting regions of the parameter space, will be located in the fluid regime.

We again consider a uniform density background and
ignore the orbital motion of the DM background with respect to the center of mass of the whole system, $v_{w}\simeq0$. Under the latter condition, we can neglect the centrifugal and Coriolis forces which may affect the density wakes and the drag forces \cite{Adams:1989,Ostriker:1992}. 
We now follow closely \cite{Kim:2007,Kim:2008} and assume that the perturbed density field over the background $\alpha=(\rho-\rho)/\rho_0$ is adiabatic and $\alpha\ll1$. At linear order, the equations of hydrodynamics imply a three-dimensional wave equation for $\alpha$:
\begin{align}\
\begin{split}
     \nabla^2\alpha-\frac{\partial^2\alpha}{c_s^2\partial^2t}=-\frac{4\pi G}{c_s^2}\rho_{\rm ext}(\mathbf{x},t),
\end{split}
\label{eq:perturber}
\end{align}
where $\rho_{\rm ext}$ denotes the mass density of the {\it perturbers}, i.e., the binary members. Cylindrical coordinates $(r,\theta,z)$ are appropriate for our case, and for simplicity we will assume that the pulsars are point masses with equal mass $M$. They are located at $(r_p,0,0)$ and $(r_p,\pi,0)$ at $t=0$. Thus, the density of the perturbers is given by:
\begin{align}
\begin{split}
     \rho_{\rm ext}(\mathbf{x},t)=&MH(t)\delta(r-r_p)\delta(z)\cdot\\
   &(\delta[r_p(\theta-\Omega t)] +\delta[r_p(\theta-\pi-\Omega t)]),
\end{split}
\end{align}
where $\Omega$ is the angular speed of the perturbers and $H(t)$ is the Heaviside step function. 

The drag force on each perturber $i$   appearing in \eqref{eq:DMbin} is given by \cite{Kim:2007,Kim:2008}:
\begin{align}
\begin{split}
  \mathbf{F}_i^{\DM}=
  \mathbf{F}^{\DM}_{i,1}+\mathbf{F}^{\DM}_{i,2}
 \end{split}
\end{align}
where the force is split in two components, one due to the perturber's own wake and the other due to the wake of the companion. Both of them can be decomposed into a radial and an azimuthal component in the following form (we remove the $i$ dependence to avoid cluttered notation)
\begin{eqnarray}
  \mathbf{F}^{\DM}_{1}&=&-\mathcal{F}(I_{1,r}\hat{\mathbf{r}}+I_{1,\theta}\hat{{\bm\theta}}),\nonumber\\
    \mathbf{F}^{\DM}_{2}&=&-\mathcal{F}(I_{2,r}\hat{\mathbf{r}}+I_{2,\theta}\hat{\bm{\theta}}),
\end{eqnarray}
with $\mathcal{F}\equiv 4\pi \rho_0(GM)^2/v_p^2$, where $v_p$ is the orbital velocity of the perturbers, and where $I_{j,r}, I_{j,\theta}$ are the dimensionless drag forces on the perturbers (due to its own wake for $j = 1$ and due to its companion's wake for $j = 2$). 
\begin{strip}
For the calculation of the period's variation, just the azimuthal components will be taken into account. We use the algebraic fits derived in \cite{Kim:2007,Kim:2008}, which are accurate within 6-16\% for all Mach numbers $\mathcal{M}$:

\begin{equation}
               I_{1,\theta}=
             \begin{cases} 0.7706\ln \left(\frac{1+\mathcal{M}}{1.0004-0.9185\mathcal{M}}\right)
               -1.4703\mathcal{M}&\mbox{if}\; \mathcal{M} < 1.0, \\  
               \ln \left(330\left(\frac{r_p}{r_{\rm min}}(\mathcal{M}-0.71)^{5.72}\mathcal{M}^{-9.58}\right)\right)& \mbox{if}\; 1.0  \leq \mathcal{M} \leq 4.4, \\ 
               \ln \left(\frac{r_p/r_{\rm min}}{0.11\mathcal{M}+1.65}\right)& \mbox{if}\; \mathcal{M}\geq 4.4, \end{cases}
\label{eq:wake_1}
 \end{equation}
\\and
\begin{align}
\begin{split}
		I_{2,\theta}=\begin{cases} \mathcal{M}^2\left[-0.022(10-\mathcal{M}){\rm tanh}(3\mathcal{M}/2)\right]& \mbox{if}\; \mathcal{M} < 2.97, \\  
		\mathcal{M}^2\left[-0.13+0.07{\rm tan}^{-1}(5\mathcal{M}-15)\right] & \mbox{if}\; \mathcal{M} \geq 2.97, \end{cases}
\end{split}
\label{eq:wake_2}
\end{align}
\end{strip}
where $r_{\rm min}$ represents the cutoff radii introduced to avoid divergences of the force integrals and it is assumed to be of the order of the characteristic size of the perturber. In order to make a proper comparison with the collisionless approach, we choose the cutoff radius such that $\ln r_p/r_{min} \sim15$, of the same order as the Coulomb logarithm $\lambda$ defined in Eq.~\eqref{df_0}. We remark that this is consistent with the logarithm of the ratio of the size of the orbit to the characteristic size of the compact objects we are dealing with, which is $\mathcal O(10$~km). For instance, if we take a perturber with a mass $M=1.3$~$M_{\odot}$, orbiting with a period of 10 days ($a\sim0.1$~au), we get $\ln (\frac{GM_p/v_p^2}{10~{\rm km}})\simeq14$. Thus, beyond $\mathcal{M} \gg 4.4$, 
 it is reasonable to assume that $I_{1,\theta}, I_{2,\theta}$ are approximately constant given the large value of $\ln r_p/r_{min}$ (see Eq.~\eqref{eq:wake_1}). 

\subsection{Numerical results: comparison with the collisionless approach}

We here compare the results in the previous section to the collisionless case of Sec.~\ref{sec:collisionless}, both for the case where the wakes' interactions are taken into account and for the case when they are neglected. Since the fluid approach we discuss was derived for equal mass systems, we concentrate in 
the case with $M_1=M_2= 1.3$~$M_{\odot}$ and focus on the
interesting range of $\sigma$ for the DDDM model, $1~{\rm km/s} \leq \sigma \leq 10~{\rm km/s}$.
As before, we consider two different values for the disk scale length, namely $z_d=10$~pc (thin disk) and $z_d=100$~pc (thick disk). Each of these cases has observational upper limits to the velocity dispersion (see Eqs.~\eqref{eq:bound_thick}$-$\eqref{eq:bound_thin}). Thus, for the thin disk, the region of interest is $\sigma\in(1,2)$~km/s, while for the thick disk it is, $\sigma\in(1,9)$~km/s.

\begin{figure}
      \centering
\includegraphics[height=7.5cm,width=7.5cm,keepaspectratio]{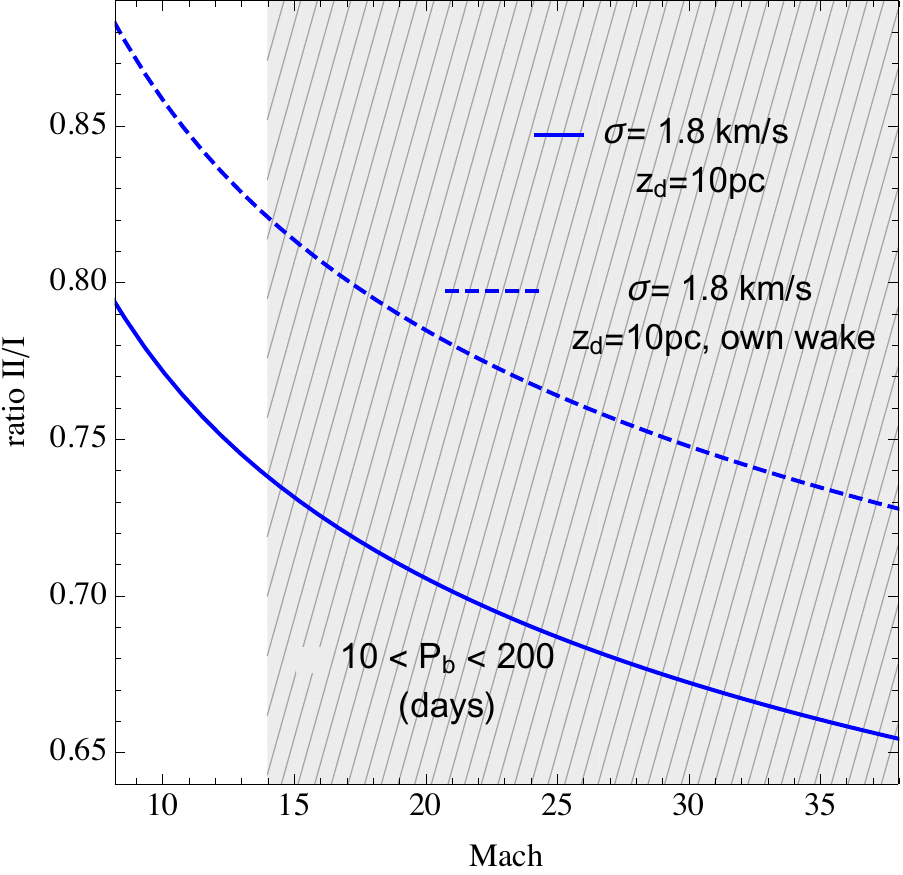}
\includegraphics[height=7.5cm,width=7.5cm,keepaspectratio]{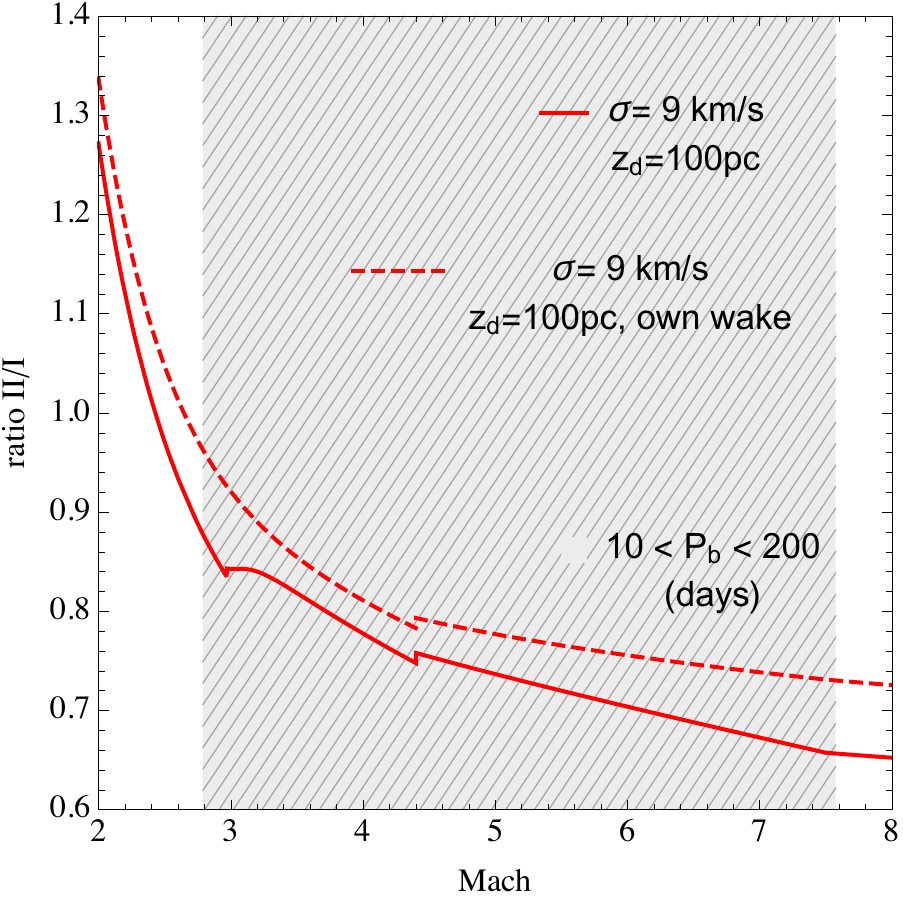}
\caption{Top (Bottom) panel: Ratios between the value of $\dot{P}^\DDDM_b$ in the fluid approach and the collisionless approach as a function of the Mach number, for a thin (thick) dark disk with $z_d=10(100)$~pc, and with a value of $\sigma$ close to current upper bounds. We show the results both including and not including the contribution of the companion wake $I_{2,\theta}$ in the collisional formulation.The binary is set to an equal mass system with $M_1=M_2=1.3$~$M_{\odot}$, $P_b=100$ days, and with coordinates in the Galaxy: $R=R_{\odot}$, $z= z_d$. The shaded area marks the range of periods of the most promising binaries shown in Fig.~\ref{fig:data_comparison}.}
\label{fig:app_comp_1}
\end{figure}

In Fig. \ref{fig:app_comp_1} we plot the ratio between the fluid and collisionless  approaches in the cases of a thin (top panel) and thick (bottom panel) disks,  choosing $\sigma$ in each case to be close to the current upper limits on a dark disk. The plot shows the dependence on the Mach number; notice that for a fixed value of $\sigma$, a high Mach number implies a smaller period.
We focus on the region $10\ \mathrm{days}\leq P_b\leq 200\ \mathrm{days}$ (shaded area). We also show the results neglecting the contribution of the companion wake ($I_{2,\theta}=0$) in the fluid approach (dashed curves). Although the relevant range of Mach numbers is different for the thick and thin disc cases (since the velocity dispersions are different), the difference between both the fluid and collisionless approaches is similar, and amounts to less than a factor of two. Also in both cases, the influence of the companion's wake is relatively minor. Notice that in the case of the thick disk, 
for $\mathcal{M}\sim2$ the drag force in the fluid approach is in fact larger than the one in the collisionless approach (although not shown in the plot this situation is more pronounced for $\mathcal{M}\sim1$). This is to be expected: for such low Mach numbers the contribution of $I_{1,\theta}$ (own wake) clearly dominates the drag force (see Fig.2 in \cite{Kim:2008}).    Near $\mathcal{M}\simeq1$ this term is much more relevant in the fluid case than in the collisionless formulation: perturbers moving at speeds near $\mathcal{M}=1$ resonantly interact with the pressure waves that they generate in the background medium \cite{Ostriker:1999}.
In the case of the thin disk (top panel of Fig.~\ref{fig:app_comp_1}), the relevant Mach numbers are quite high $\mathcal{M}\gtrsim15$,
which makes the influence of the companion more relevant, albeit still small. The final result is roughly a $10-15\%$ reduction over the effect of
the principal (own) wake.

Therefore, quantitatively, the difference between the two approaches depends on the details of the specific system in consideration and the specific features of the dark disk. However, for the range of periods $10$~days$\,\leq P_b\leq200$ ~days (marked with a shaded area in Fig.~\ref{fig:app_comp_1}) where we find the most promising binary pulsars (those shown in Fig.~\ref{fig:data_comparison}), the predicted orbital decay is of the same order of magnitude for both approaches, while the interaction of the pair of wakes seems to play only a minor role.

We note that the bump seen in the bottom panel of Fig.~\ref{fig:app_comp_1} in the range $2.97<\mathcal{M}<4.4$ is related to the overlapping of the two wake tails, which enhances the drag force in the perturber (see Fig. 2 of \cite{Kim:2008}). The sharp features around the bump are exacerbated by artificial discontinuities in the fitting functions for  $I_{1,\theta}$ and $I_{2,\theta}$ in Eqs.~\eqref{eq:wake_1}$-$\eqref{eq:wake_2}. 

In summary, the results of the comparison between the different approaches suggest that the main (own) wake of each member of the binary dominates the change in the orbital period, with the effect of the companion wake leading only to a $10-15\%$ reduction in the expected signal. Although the latter is strictly valid only in the fluid case, our results bring confidence that the potential constraining power of binary pulsars suggested in Section~\ref{sec:collisionless} (particularly Fig.~\ref{fig:data_comparison} for the {\it absolute} value of $\dot P_b$) in the DDDM model, is expected to remain valid even when the interaction between wakes is taken into account. We expect to substantiate this hope with explicit calculations in the future.

\section{Summary and Conclusions}\label{sec:summary}

We study the signature of a dark matter (DM) disk in the orbital change of a binary pulsar due to the induced gravitational force generated by the DM wakes formed as the binary moves through the DM background. In particular, we concentrate on the Double-Disk-Dark-Matter  (DDDM) scenario introduced in \cite{Fan:2013,Fan:2013PRL}, which exhibits all the properties needed to enhance this effect, originally introduced by \cite{Pani:2015} in the context of the traditional non-dissipative Cold Dark Matter (CDM) scenario. Contrary to the standard CDM halo, if a subdominant DM component is made of dissipative DM, it would form a rotationally supported flat disk with larger density and much lower velocity dispersion than the CDM triaxial halo, and which is also approximately co-rotating with the baryonic disk. We find that all these features enhance by many orders of magnitude the impact of DM in the orbital change of pulsars, relative to the standard case studied by \cite{Pani:2015}. 
It is important to remark that other DM models with dissipative properties may also generate disks, and hence our results might be of interests in those models as well, see e.g. \cite{DAmico:2017lqj,Foot:2014uba,Arkani-Hamed:2016rle,Buckley:2017ttd}.

We model the dark disk as an isothermal sheet (Eq.~\eqref{eq_rho}) with a radial scale length $R_d=3$~kpc, equal to that of the baryonic disk, and explore the cases where the vertical scale length is set to $z_d=10$~pc (thin disc) and $z_d=100$~pc (thick disc). This range covers the  interesting cases discussed in the literature. To maximize the impact of these scenarios, we set the density (velocity dispersion) to the largest possible values according to updated stellar kinematics constraints \cite{Kramer:2016} (see Eqs.~\eqref{eq:bound_thick}$-$\eqref{eq:bound_thin})

The impact of the presence of a dark disk in the orbital period of a binary pulsar is investigated with two approaches:
\begin{itemize}
\item[i)] a {\it collisionless approach} where DM particles are treated as collisionless, and where each pulsar is influenced {\it only} by its own induced wake (see Section~\ref{sec:collisionless}). This approach was introduced in \cite{Pani:2015}, with the important difference that we explore the regime suitable for a dark disk: $v_w\ll\sigma\ll v_{\rm orb}$, where $v_w$ is the relative velocity between the center of mass of the binary and the DM background ($v_w\sim0$ is expected for some systems for a co-rotating disk), $\sigma$ is the velocity dispersion and $v_{\rm orb}$ is the orbital velocity of the binary. In contrast, the regime studied in \cite{Pani:2015} is appropriate for a CDM halo where $v_w\gtrsim\sigma\gtrsim v_{\rm orb}$.
\item[ii)] a {\it fluid approach} where DM particles are assumed to be in the fluid regime and for a binary of {\it equal masses} where the interaction of the wake pair is fully considered \cite{Kim:2007,Kim:2008}. This approach is expected to be valid when the Knudsen number $Kn\ll1$ (where the mean free path is given by the Rutherford scattering of the charged particles, and the characteristic size of interest is the size of the binary's orbit), which is expected to hold in some regions of the relevant parameter space of the DDDM scenario (see Section~\ref{sec:fluid}). 
\end{itemize}

The consideration of both approaches allows us to give a broad perspective of the problem of interest. In particular, since $v_{\rm orb}/\sigma\gtrsim 1$, the motion is supersonic, which suggests that the wake of the companion could be significant (something that is ignored in the collisionless approach so far developed). 
For this reason, the case with $Kn \ll 1$ is important, since it allows us to explore whether the main effect is well captured by the calculations that ignore the wakes' interactions.  The intermediate regime, that of $Kn\sim 1$, is also of interest and remains entirely unexplored. Our intention is to develop a general approach that covers all regimes in the near future. In particular, we highlight the importance to develop, in the collisional approach, a treatment for unequal-mass perturbers, which is the case for most of the observed systems. In \cite{Kim:2008}, in fact, the authors assumed identical orbits for both perturbers, and this assumption would fail soon for unequal masses. Nevertheless, in the present paper our goal is to show the promising aspects of binary pulsars as probes of the DDDM scenario, and to present general expectations based on a hybrid analysis that combines the results from the methods currently available. 

Using the first (collisionless) approach, we estimate the orbital period variation $\dot{P}^\DDDM_b$ due to a hypothetical DDDM disk for many binary pulsars within that dark disk (see Table~\ref{table:systems}). In a couple of cases, our theoretical predictions are relatively close (within a factor of $\sim100$) to current experimental bounds/values of $\dot{P_b}$ (see Fig.~\ref{fig:data_comparison}). These systems are very promising for future follow ups, although we used the approximation $v_w\lesssim \sigma$ which is likely not satisfied in these cases given their transverse velocities. Still, given the capacity of future surveys, such as SKA \cite{Kramer:2015bea}, to detect new (up to ${\cal O}(10^3)$) systems and get timing measurements with enough precision, we may soon be able to find binary pulsars that satisfy all the required properties to probe the DDDM scenario.

Using the second (fluid) approach, we have estimated the impact of the interacting pair of wakes in the orbit of the binary. We find that the companion's wake has only a minor role ($10-15\%$ effect). Furthermore, we also show that, for the most interesting binary candidates, the induced $\dot{P}^\DDDM_b$ is of the same order for the three cases we studied: fluid, fluid but neglecting wakes' interactions and collisionless neglecting the
wakes' interactions, see Fig.~\ref{fig:app_comp_1}.  In particular, when the dark disk gets thicker ($\sigma$ increases), the Mach number becomes smaller and the companion's wake is less relevant. 
Overall, our results suggest that the expectation of a change in the period of the orbit of binary pulsars of the order of magnitude estimated using the first (collisionless) approach, will hold even if the wakes' interaction were taken into account into a full collisionless approach (something that has not been developed so far).

Thus, we conclude that the precise observations of binary pulsars may probe the scenarios of dark disks with properties close to those of allowed DDDM scenarios if the precision of the most promising candidates is improved  or if  future surveys discover new systems close to the galactic plane and with long enough periods. 
As we discussed, we leave for the future a more detailed analysis of the case $Kn \sim 1$ and a more complete comparison with data.

\section*{Acknowledgments}

DB is grateful to Matthew McCullough, Paolo Pani and Alberto Sesana for discussions.
We are grateful to Lijing Shao for comments on the draft. {This work was originally inspired by a research visit made by JZ to CERN in the context of the CERN-CKC TH institute during the Summer of 2016, which was funded by CERN and the University of Iceland.
JZ acknowledges support by a Grant of Excellence from the
Icelandic Research Fund (grant number $173929-051$). AC acknowledges support by the European project H2020- MSCA-ITN-2015//674896-ELUSIVES.

\appendix

\section{Orbital parameters and predictions for a broad sample of binary pulsars}\label{ap:table}

\begin{strip}

\label{Table_full}
\centering	
\begin{tabular}{lcccccc}
\hline
 Name& $P_b$(days) & $z_{\rm sys}$ (kpc)& $v_{\rm orb}({\rm km/s})$ & $\dot{P}_b^\DDDM$ & $(\dot{P_b})_{\rm lim}$ & Ref.\\
 \hline
 J1751-2857   & 110.7    &-0.02& $61.7$ & $-2.1\cdot10^{-13}$ & $1.8\cdot10^{-11}$ & \cite{Desvignes:2016}\\ 
  J1857+0943   & 12.3    &0.06& $128.4$ & $-9.76\cdot10^{-16}$ & $1.2\cdot10^{-13}$ & \cite{Desvignes:2016}\\ 
 J0621+1002   & 8.3    &-0.01& $146.4$ & $-5.79\cdot10^{-16}$ & $10^{-7}$ & \cite{Desvignes:2016}\\
 J1012-4235&   37.97  & 0.07   & 88.2&$-7.06\cdot10^{-15}$&$10^{-3}$ & \cite{Camilo:2015}\\
 J1017-7156&6.51 & -0.06& 158.7 &$-4.71\cdot10^{-16}$&$10^{-6}$ & \cite{Ng:2014} \\
 J1125-5825&76.40 &0.08 & 69.9 &$-2.13\cdot10^{-14}$&$10^{-8}$ & \cite{Ng:2014}\\
J1125-6014&8.75 &0.02 & 143.9 &$-9.88\cdot10^{-16}$&$10^{-8}$ & \cite{Lorimer:2006}\\
 B1259-63&  1236.72&-0.04 & 27.6 &$-6.07\cdot10^{-12}$&$\reff$ $1.47\cdot10^{-8}$ & \cite{Shannon:2014}\\
 J1420-5625& 40.29  & 0.01 & 86.5 &$-2.18\cdot10^{-14}$&$10^{-6}$ & \cite{Hobbs:2004}\\
 J1454-5846& 12.42  & 0.02& 128.0 &$-6.49\cdot10^{-16}$&$10^{-8}$ & \cite{Camilo:2001}\\
J1543-5149& 8.06  & 0.05& 147.9 &$-6.47\cdot10^{-16}$&$10^{-9}$ & \cite{Ng:2014}\\
J1638-4725& 1940.9 & -0.02& 23.8 &$-6.01\cdot10^{-11}$&$10^{-1}$ & \cite{Lorimer:2006}\\
J1727-2946& 40.31  &0.01 & 86.5 &$-6.28\cdot10^{-15}$&$10^{-8}$ & \cite{Lorimer:2015}\\
J1740-3052& 231.03  & -0.01& 48.3 &$-4.40\cdot10^{-12}$&$\reff$ $3\cdot10^{-9}$ & \cite{Madsen:2012}\\
J1750-2536& 17.14  &0.04 & 115.0 &$-8.21\cdot10^{-16}$&$10^{-6}$ & \cite{Knispel:2013}\\ 
J1753-2240& 13.64  &0.09 & 124.1 &$-2.91\cdot10^{-16}$&$10^{-7}$ & \cite{Keith:2009}\\
J1755-25& 9.69  & -0.03& 139.0 &$-3.41\cdot10^{-16}$&$10^{-5}$ & \cite{Ng:2015}\\
J1810-2005&15.01 & -0.03& 120.2 &$-1.30\cdot10^{-15}$&$10^{-8}$ & \cite{Janssen:2010}\\
J1811-1736& 18.78  &0.03 & 11.5 &$-4.20\cdot10^{-16}$&$10^{-7}$ & \cite{Corongiu:2007}\\
B1820-11& 357.76  & 0.09& 41.8  &$-4.78\cdot10^{-14}$&$10^{-5}$ & \cite{Hobbs:2004}\\
J1840-0643& 937.1  &-0.05 & 30.3  &$-2.87\cdot10^{-13}$&$10^{-1}$ & \cite{Knispel:2013}\\
J1850+0124& 84.94  & 0.06& 67.4 &$-1.84\cdot10^{-13}$&$10^{-6}$ & \cite{Crawford:2012}\\
J1903+0327& 95.17  &-0.03 & 64.9 &$-9.42\cdot10^{-15}$&$\reff$ $-33\cdot10^{-12}$ & \cite{Freire:2011} \\
J1910+1256& 58.47  & 0.05& 76.4 &$-1.41\cdot10^{-14}$&$10^{-9}$ & \cite{Fonseca:2016}\\
J1935+1726& 90.76 & -0.06& 66.0 &$-1.46\cdot10^{-14}$&$10^{-5}$ & \cite{Lorimer:2013}\\
B1953+29& 117.35  & 0.05& 60.6 &$-7.80\cdot10^{-15}$&$10^{-8}$ & \cite{Fonseca:2016}\\
J1737-0811& 79.51 & 0.04& 68.9 &$-3.52\cdot10^{-14}$&$10^{-6}$ & \cite{Boyles:2013}\\
J1943+2210& 8.31 & -0.09& 146.4 &$-5.53\cdot10^{-17}$&$10^{-8}$ & \cite{Scholz:2015}\\
\hline
   
\end{tabular}


\captionof{table}{A selection of binary pulsars near the galactic plane that are potential targets to constraint the DDDM model. The columns are (1) Name. (2) Binary period. (3) Distance to the galactic plane. (4) Orbital velocity of the binary. (5) Maximal time variation in the binary period due to the presence of a dark disk. (6) Observational upper bound on $\dot{P_b}$; those few systems with the tag ``$\reff$'' have a measurement on $\dot{P_b}$. (7) Reference. }
\label{table:systems}

\end{strip}

\bibliography{lit}
\bibliographystyle{elsarticle-harv}

\end{document}